\documentclass[apjl]{emulateapj}
\usepackage{amsfonts,amsmath,graphicx,natbib,apjfonts,lscape}

\def\grb{GRB\,091127}

\def\cfa{1}
\def\clay{2}
\def\lbnl{3}
\def\ucb{4}
\def\oxf{5}
\def\psu{6}
\def\gem{7}

\begin{document}

\title{The Spectroscopic Classification and Explosion Properties of
SN\,2009nz Associated with GRB\,091127 at $z=0.490$}

\author{
E.~Berger\altaffilmark{\cfa},
R.~Chornock\altaffilmark{\cfa},
T.~R.~Holmes\altaffilmark{\cfa},
R.~J.~Foley\altaffilmark{\cfa,}\altaffilmark{\clay},
A.~Cucchiara\altaffilmark{\lbnl,}\altaffilmark{\ucb},
C.~Wolf\altaffilmark{\oxf},
Ph.~Podsiadlowski\altaffilmark{\oxf},
D.~B.~Fox\altaffilmark{\psu},
and K.~C.~Roth\altaffilmark{\gem}
}

\altaffiltext{\cfa}{Harvard-Smithsonian Center for Astrophysics, 60
Garden Street, Cambridge, MA 02138}

\altaffiltext{\clay}{Clay Fellow}

\altaffiltext{\lbnl}{Lawrence Berkeley National Laboratory, M.S. 50-F,
1 Cyclotron Road, Berkeley, CA 94720}

\altaffiltext{\ucb}{Department of Astronomy, 601 Campbell Hall,
University of California, Berkeley, CA 94720-3411}

\altaffiltext{\oxf}{Department of Astrophysics, Denys Wilkinson
Building, University of Oxford, Keble Road, Oxford OX1 3RH}

\altaffiltext{\psu}{Department of Astronomy and Astrophysics,
Pennsylvania State University, 525 Davey Laboratory, University Park,
PA 16802}

\altaffiltext{\gem}{Gemini Observatory, 670 N. Aohoku Place Hilo, HI
96720}

\begin{abstract} We present spectroscopic observations of GRB\,091127
($z=0.490$) at the peak of the putative associated supernova
SN\,2009nz.  Subtracting a late-time spectrum of the host galaxy, we
isolate the contribution of SN\,2009nz and uncover broad features
typical of nearby GRB-SNe.  This establishes unambiguously that
GRB\,091127 was accompanied by a broad-lined Type Ic SN, and links a
cosmological long burst with a standard energy release ($E_{\rm
\gamma,iso}\approx 1.1\times 10^{52}$ erg) to a massive star
progenitor.  The spectrum of SN\,2009nz closely resembles that of
SN\,2006aj, with SN\,2003dh also providing an acceptable match, but
has significantly narrower features than SNe 1998bw and 2010bh,
indicative of a lower expansion velocity.  The photospheric velocity
inferred from the \ion{Si}{2}$\lambda 6355$ absorption feature,
$v_{\rm ph}\approx 17,000$ km s$^{-1}$, is indeed closer to that of
SNe 2006aj and 2003dh than to the other GRB-SNe.  Combining the
measured velocity with the light curve peak brightness and width, we
estimate the following explosion parameters: $M_{\rm Ni}\sim 0.35$
M$_\odot$, $E_{K}\sim 2.3\times 10^{51}$ erg, and $M_{\rm ej}\sim 1.4$
M$_\odot$, similar to those of SN\,2006aj.  These properties indicate
that SN\,2009nz follows a trend of lower $M_{\rm Ni}$ for GRB-SNe with
lower $E_{K}$ and $M_{\rm ej}$.  Equally important, since GRB\,091127
is a typical cosmological burst, the similarity of SN\,2009nz to
SN\,2006aj either casts doubt on the claim that XRF\,060218/SN\,2006aj
was powered by a neutron star, or indicates that the nature of the
central engine is encoded in the SN properties but not in the prompt
emission.  Future spectra of GRB-SNe at $z\gtrsim 0.3$, including
proper subtraction of the host galaxy contribution, will shed light on
the full dispersion of SN properties for standard long GRBs, on the
relation between SNe associated with sub-energetic and standard GRBs,
and on a potential dispersion in the associated SN types.
\end{abstract}


\section{Introduction}
\label{sec:into}

The association of long-duration gamma-ray bursts (GRBs) with Type Ic
supernovae (SNe) provides the most direct evidence for massive
stripped-envelope (Wolf-Rayet) stars as the progenitors of long GRBs
(e.g., \citealt{wb06}).  This connection was first intimated by the
discovery of the unusually luminous and broad-lined Type Ic SN\,1998bw
($d=40$ Mpc) in spatial and temporal coincidence with GRB\,980425
\citep{gvv+98}.  It was further supported by the detection of
late-time photometric ``bumps'' in the optical light curves of several
GRBs at $z\lesssim 1$ that resembled the optical light curve of
SN\,1998bw (e.g., \citealt{bkd+99,lcg+01,bkp+02,skf+05,skp+06}),
although alternative explanations for these bumps have been proposed
\citep{eb00,wd00}.  Subsequently, unambiguous spectroscopic
identifications were obtained for four nearby GRBs:
GRB\,030329/SN\,2003dh ($z=0.169$; \citealt{hsm+03,smg+03}),
GRB\,031203/SN\,2003lw ($z=0.105$; \citealt{mtc+04}),
XRF\,060218/SN\,2006aj ($z=0.033$; \citealt{msg+06,pmm+06}), and
GRB\,100316D/SN\,2010bh ($z=0.059$; \citealt{cbl+10}).  All four
events are broad-lined Type Ic SNe, similar to SN\,1998bw.  However,
only GRB\,030329 is typical of the cosmological long GRB population in
terms of its energy scale \citep{bkp+03}, while the remaining three
events are all sub-energetic in $\gamma$-rays, have no clear optical
afterglows, exhibit only mildly relativistic velocities, and appear to
be quasi-spherical (the so-called sub-energetic GRBs;
\citealt{skb+04,skn+06}).

As a result, there is a clear impetus to obtain spectroscopic
observations of GRB-SNe at $z\gtrsim 0.3$ to bolster the GRB-SN
connection for the cosmological population, as well as to map the
range of GRB-SN properties (i.e., velocity, kinetic energy, ejecta
mass, $^{56}{\rm Ni}$ mass), and to compare the properties of the SNe
associated with standard and sub-energetic GRBs.  The challenge of
obtaining such spectroscopic observations is twofold.  First, the
associated SNe are faint, with a peak optical apparent
brightness\footnotemark\footnotetext{This is due in part to
line-blanketing in the rest-frame UV, which is redshifted into the
optical band at $z\gtrsim 0.5$.} of $i_{\rm AB}\approx 22.5$ mag at
$z\sim 0.5$ and $\approx 24.5$ mag at $z\sim 1$.  Second, at these
redshifts the brightness of the SNe and the compact host
galaxies\footnotemark\footnotetext{The typical half-light radii of
long GRB host galaxies at $z\gtrsim 0.3$ are $\lesssim 0.5''$
\citep{bkd02,wbp07}.} are similar, therefore requiring a careful
subtraction of the host spectrum to robustly reveal and properly
measure the SN features.

To date, spectroscopic observations timed to coincide with potential
GRB-SNe at $z\sim 0.3-1$ have been obtained in five cases.
\citet{dmb+03} obtained a spectrum of GRB\,021211 ($z=1.006$) during
the peak of an associated bump (designated SN\,2002lt).  The putative
SN and host galaxy had a comparable brightness of $R\approx 25.2$ mag,
but the host contribution was not subtracted from the spectrum, which
appears to exhibit a broad minimum interpreted as \ion{Ca}{2} H\&K
blue-shifted by about $15,000$ km s$^{-1}$.  \citet{dmb+03} claim a
close match with the normal Type Ic SN\,1994I, but a poor match with
SN\,1998bw; however, inspection of their Figure 2 shows that the
peculiar Type Ia SN\,1991bg is an equally good match to the noisy
spectrum of the putative SN\,2002lt.  \citet{gsw+03} obtained a
spectrum of GRB\,011121 ($z=0.362$) coincident with the peak of an
associated bump (designated SN\,2001ke) when the optical
brightness\footnotemark\footnotetext{It is unclear from \citet{gsw+03}
what fraction of the flux within the $1''$ slit is contributed by the
extended host galaxy.  The host contribution was not subtracted from
the spectrum.} was $R\approx 23$ mag.  Unlike SN\,1998bw, the
resulting spectrum is flat at $\lambda_{\rm rest} \lesssim 5100$ \AA,
which \citet{gsw+03} claim as evidence for a Type IIn event.  However,
the large, and uncertain, extinction along this line-of-sight may
influence this result.  A spectrum of XRF\,020903 ($z=0.251$) was
obtained by \citet{skf+05} during a late flattening in the light
curve.  After subtraction of a model starburst galaxy template, the
residual spectrum exhibits an overall shape similar to SN\,1998bw.
\citet{dmb+06} obtained a spectrum of GRB\,050525A ($z=0.606$) about 2
weeks after the peak of the associated bump (designated SN\,2005nc),
with equal contribution from the putative SN and host galaxy (each
with $R\approx 25.2$ mag).  After subtraction of a model host
spectrum, the residual spectrum exhibits a broad bump at $\lambda_{\rm
rest}\approx 5100$ \AA, with a decline at longer wavelengths, broadly
resembling the spectrum of SN\,1998bw.  \citet{ssf+11} obtained
spectra of GRB\,101219B ($z=0.552$ based on weak \ion{Mg}{2}
absorption) about 16 and 37 d after the burst and find broad
undulations that resemble SN\,1998bw and distinct from an earlier
featureless spectrum.  They do not directly account for the spectral
shape of the host galaxy, assuming instead a flat
spectrum\footnotemark\footnotetext{This assumption was made despite
the fact that late-time photometric observations indicate a lower flux
in the observed $g$-band compared to $r$- and $i$-band.}.  In none of
the 5 cases was an attempt made to extract any SN properties beyond
addressing whether or not they resembled SN\,1998bw.

Here we present spectroscopic observations of GRB\,091127 at
$z=0.490$, timed to coinciding with the peak of a photometric bump
designated SN\,2009nz \citep{cbp+10}.  The bump light curve resembles
SN\,1998bw, albeit dimmer by about 0.1 mag and with a somewhat earlier
peak and narrower light curve; it is brighter than SN\,2006aj by about
0.25 mag, with a later peak and broader light curve \citep{cbp+10}.
At the peak of the bump, the SN brightness is $I\approx 22.3$ mag, the
afterglow is estimated to be $I\approx 24.2$ mag, and the host galaxy
has $I\approx 22.5$ mag (i.e., the relative contributions are about
$50\%$, $8\%$, and $42\%$, respectively).  Thus, as in the previous
cases of $z\gtrsim 0.3$ bumps, the host and putative supernova
contribute a similar amount of flux, thereby requiring a careful
subtraction of the host galaxy spectrum.  Here we undertake this
procedure using a spectrum of the host galaxy obtained about 1 yr
after the burst (\S\ref{sec:obs}).  Subtraction of the host
contribution cleanly isolates the SN spectrum and reveals broad
undulations that clearly resemble previous GRB-SNe, in particular
SN\,2006aj and potentially SN\,2003dh, but with narrower features than
SN\,1998bw and SN\,2010bh (\S\ref{sec:results}).  This result
demonstrates that the photometric bump is indeed a SN, and allows us
to extract the explosion properties of SN\,2009nz
(\S\ref{sec:results}).  We compare these properties to the
well-studied nearby GRB-SNe to explore the dispersion of SN properties
for standard and sub-energetic GRBs (\S\ref{sec:conc}).

\section{Spectroscopic Observations}
\label{sec:obs}

\subsection{``Bump'' Spectrum}

We observed GRB\,091127 ($z=0.490$; \citealt{gcn10202}) with the
Gemini Multi-Object Spectrograph (GMOS) on the Gemini-North 8-m
telescope (program GN-2009B-Q-28, PI: Fox) on 2009 December 22.27 UT.
The spectrum was timed to coincide with the peak of an associated
SN\,1998bw-like event at $z=0.490$.  Two dithered 900 s exposure pairs
were obtained with the R400 grating at central wavelengths of 8000 and
8050 \AA\ in $0.65''$ seeing.  The resulting combined spectrum covers
$5900-9500$ \AA\ at a resolution of about 7 \AA.  An OG515
order-blocking filter was used to prevent second-order contamination
in the red part of the spectrum.

The spectra were reduced, combined, and extracted using standard
procedures in IRAF, while flux calibration and correction for telluric
absorption were performed using custom IDL routines with archival
observations of the standard stars BD+28\,4211 and Feige\,34.  The
mean airmass of the observation was 1.3 and the $1''$ slit was aligned
at the parallactic angle, so the relative spectral shape is reliable.
We determine the overall flux scale of the spectrum by integrating
over the $r$-band filter response function and comparing to photometry
of the target in the GMOS $r$-band acquisition image within the same
aperture size ($r=22.66\pm 0.06$ mag calibrated relative to follow up
observations from Magellan -- see \S\ref{sec:magellan}).

\subsection{Host Galaxy Spectrum}
\label{sec:magellan}

Spectroscopic and photometric observations of the host galaxy of \grb\
were obtained with the Inamori Magellan Areal Camera and Spectrograph
(IMACS) mounted on the Magellan/Baade 6.5-m telescope on the nights of
2010 November 13 and 14 UT.  We obtained two 2400 s exposures with the
200 l/mm grism using a $0.9''$ slit and a 4950 \AA\ order blocking
filter to avoid second-order contamination in the red part of the
spectrum.  The seeing during the observations was $0.6''$,
well-matched to the conditions of the GMOS spectrum.  The spectra were
processed using standard procedures in IRAF, and the wavelength
calibration was performed using HeNeAr arc lamps.  The resulting
wavelength coverage is $5000-9400$ \AA, with a resolution of about 4.7
\AA.  Flux calibration was performed using observations of the
spectrophotometric standard star Feige\,110.  We confirmed the overall
flux calibration and shape by integrating the spectrum over the $r$-
and $i$-band filter response functions ($r\approx 23.5$ mag, $i\approx
23.3$ mag) compared to photometric measurements of the host within the
same aperture size ($r\approx 23.5$ mag, $i\approx 23.2$ mag).

The photometric observations were obtained in the $griz$ filters.  The
data were processed using standard procedures in IRAF, and photometric
calibration was performed using observations of the southern standard
star field E3A.  The resulting flux measurements (in a $1.5''$ radius
aperture) are provided in Table~\ref{tab:photom}.  We note that our
$i$-band measurement is in excellent agreement with the $I$-band host
flux given in \citet{cbp+10}.

\section{The Spectrum of SN\,2009nz and a Comparison to Nearby
GRB-SNe}
\label{sec:results}

\begin{figure}
\epsscale{1}
\plotone{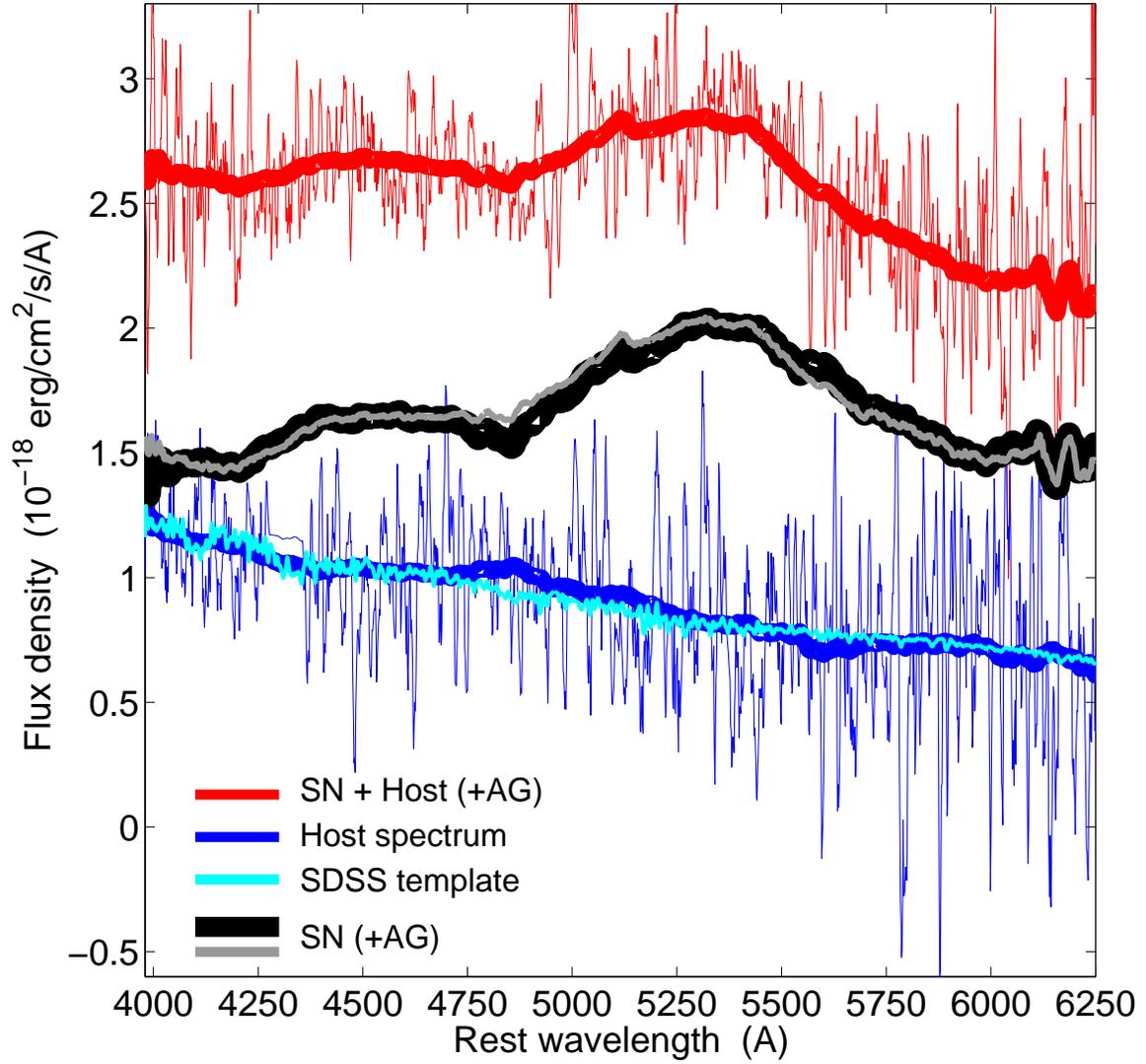}
\caption{Spectra of the GRB\,091127/SN\,2009nz ``bump'' (red) and of
the host galaxy (blue).  The thick lines are smoothed with a 100 \AA\
boxcar.  The subtracted spectrum (with about $15\%$ contribution from
the featureless afterglow: \citealt{cbp+10}) is shown as a thick black
line.  The undulations typical of a broad-lined GRB-SN are clearly
seen, with prominent absorption features at $\lambda_{\rm rest}\approx
4200$, $4750$, and $6000$ \AA.  We also perform a subtraction relative
to a template blue galaxy spectrum from the SDSS (cyan; emission lines
clipped), and find that the resulting SN spectrum (gray) is nearly 
identical to the spectrum from actual host subtraction.
\label{fig:spec}} 
\end{figure}

The spectra at the peak of the photometric bump and of the host galaxy
alone are shown in Figure~\ref{fig:spec}.  The bump spectrum
(supernova+host) exhibits broad undulations with a peak at
$\lambda_{\rm rest}\approx 5300$ \AA, and a possible second peak at
$\lambda_{\rm rest}\approx 4500$ \AA.  These features are reminiscent
of previous GRB-SNe, but it is critical to ensure that they do not
result from host galaxy contamination, since the host contributes
about half of the total flux.  To remove any possible contamination,
and to cleanly isolate the SN features, we subtract the host spectrum
from the bump spectrum and find that the features are robustly
detected.  The isolated spectrum of the transient exhibits three
absorption troughs centered at rest wavelengths of about 4200, 4750,
and 6000 \AA, closely matching the spectra of nearby GRB-SNe.  The
4200 and 4750 \AA\ features are generally identified as blends of
\ion{Fe}{2}/\ion{Ti}{2} and \ion{Fe}{2}/\ion{Fe}{3}, respectively,
while the 6000 \AA\ feature is identified as \ion{Si}{2}$\lambda
6355$.  We confirmed these results by subtracting a scaled blue galaxy
template spectrum obtained from the Sloan Digital Sky Survey
cross-correlation library\footnotemark\footnotetext{{\tt
http://www.sdss.org/dr7/algorithms/spectemplates/}}
(Figure~\ref{fig:spec}).  The galaxy spectral shape is essentially
identical to our measurements, and the resulting isolated SN spectrum
reveals the same features detected in the subtraction relative to the
Magellan host spectrum.  In the analysis below we use the SN spectrum
from the actual host subtraction, but note that the results are
unchanged for the spectrum from template subtraction.

\begin{figure}
\epsscale{0.8}
\plotone{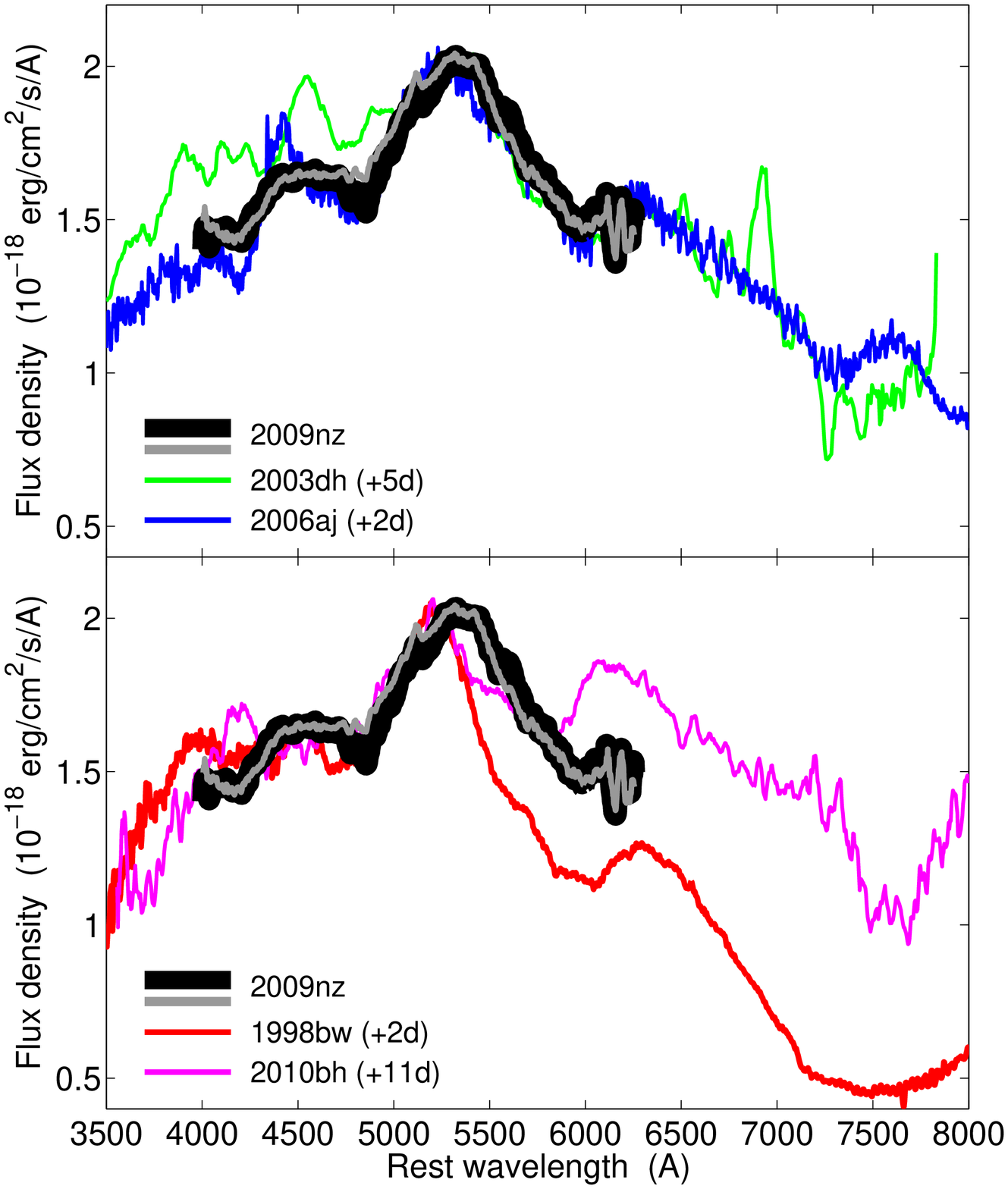}
\caption{Spectrum of SN\,2009nz (black; with about $15\%$ contribution
from the featureless afterglow: \citealt{cbp+10}) compared to spectra
of well-studied nearby GRB-SNe on a comparable timescale (the times in
parentheses are relative to the peak of the light curve).  Top:
SN\,2003dh \citep{mgs+03} and SN\,2006aj \citep{msg+06}; Bottom:
SN\,1998bw \citep{pcd+01} and SN\,2010bh \citep{cbl+10}.  The
comparison spectra have been normalized to the same flux at the peak
of the $\lambda_{\rm rest}\approx 5300$ \AA\ feature.  The spectrum of
SN\,2009nz closely resembles the previous GRB-SN, with the best
overall match provided by SN\,2006aj.  SN\,2003dh also provides a
reasonable match in terms of the width of the features, while SNe
1998bw and 2010bh have broader features indicative of larger expansion
velocities.  The spectrum from a template galaxy subtraction (gray) is
also shown.
\label{fig:comp1}} 
\end{figure}

We compare our SN\,2009nz spectrum at $\delta t_{\rm rest}=16.3$ d
(corresponding to $\approx 1.5$ d past maximum light;
\citealt{cbp+10}) to the spectra of the nearby GRB-SNe 1998bw, 2003dh,
2006aj, and 2010bh on a similar timescale (Figure~\ref{fig:comp1}).
Overall, the spectrum of SN\,2009nz is remarkably similar to the
nearby GRB-SNe, exhibiting the same broad features that are a hallmark
of large expansion velocities.  This unambiguously identifies the
photometric bump associated with \grb\ as a supernova, and
demonstrates that SN\,2009nz is also a broad-lined event,
likely\footnotemark\footnotetext{Although our spectrum does not cover
H$\alpha$, we do not see an obvious feature corresponding to H$\beta$
at an expected blue-shift of $\sim 10,000-30,000$ km s$^{-1}$.
Furthermore, the overall similarity to the nearby well-studied
GRB-SNe, which are all of Type Ic, suggests that SN\,2009nz shares
this designation.} of Type Ic.  We do not find a prominent
\ion{He}{1}$\lambda 5876$ absorption feature as expected for a Type Ib
SN.  Finally, a comparison to the normal Type Ic SN\,1994I and the
normal Type Ib SN\,1998dt reveals a poor match due to their
significantly narrower features.

\begin{figure}
\epsscale{1}
\plotone{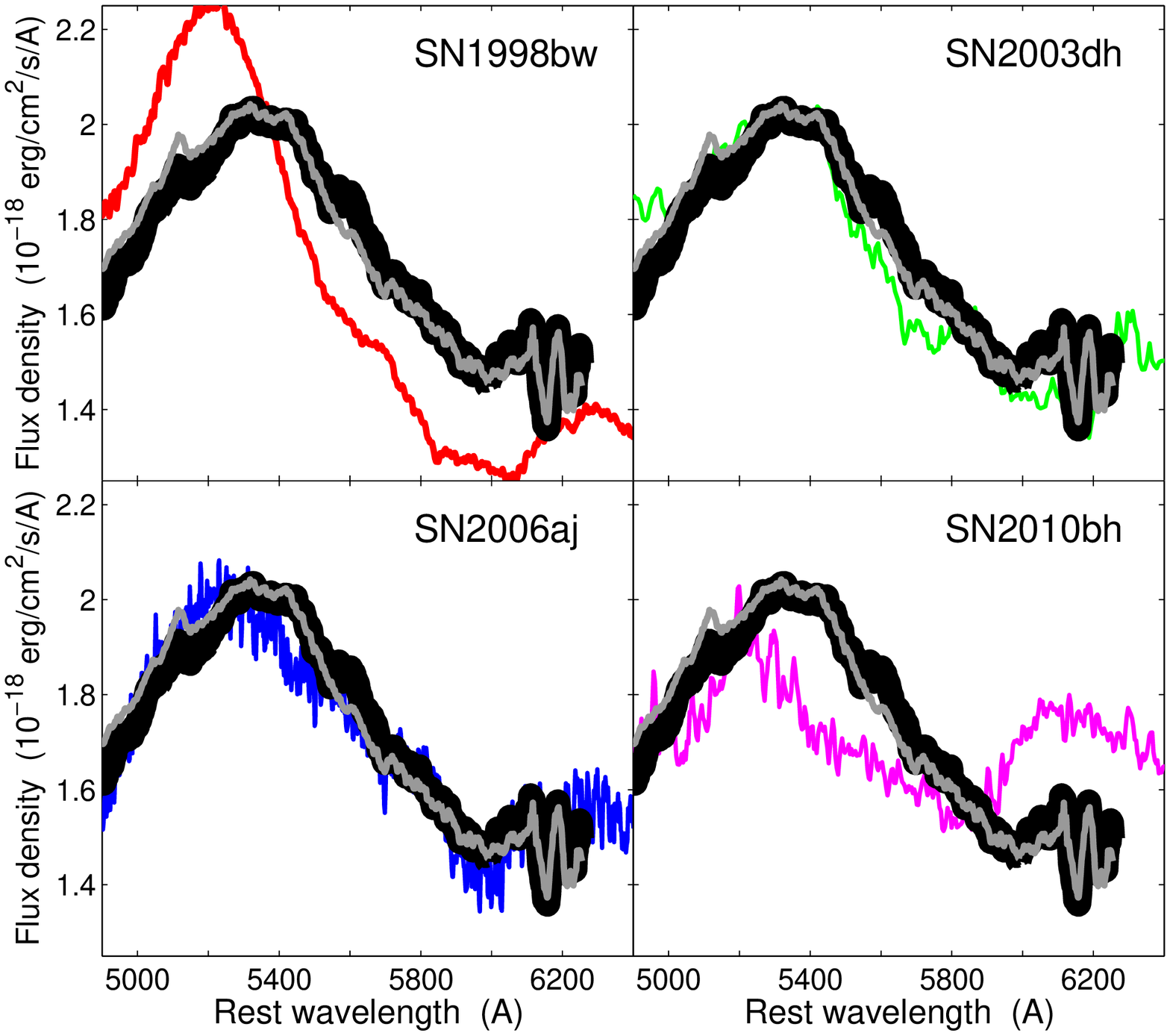}
\caption{Spectrum of SN\,2009nz near the \ion{Si}{2}$\lambda 6355$
absorption feature (black; with about $15\%$ contribution from the
featureless afterglow: \citealt{cbp+10}) compared to spectra of
well-studied nearby GRB-SNe on a comparable timescale (SN\,1998bw:
\citealt{pcd+01}; SN\,2003dh: \citealt{mgs+03}; SN\,2006aj:
\citealt{msg+06}; SN\,2010bh: \citealt{cbl+10}).  The location of the
absorption feature indicates a photospheric velocity of about $17,000$
km s$^{-1}$ for SN\,2009nz.  SNe 2006aj and 2003dh provide the closest
match in terms of the location and width of the \ion{Si}{2}$\lambda
6355$ feature, indicating a similar expansion velocity.  SNe 1998bw
and 2010bh exhibit broader features indicative of larger expansion
velocities.  The spectrum from a template galaxy subtraction (gray) is
also shown and leads to identical results.
\label{fig:comp3}} 
\end{figure}

Among the sample of nearby GRB-SNe the best overall match to
SN\,2009nz is provided by SN\,2006aj, in terms of both the location of
the absorption minima and the width of the features
(Figure~\ref{fig:comp1}).  SN\,2003dh also provides a reasonable
match, but SNe 1998bw and 2010bh have significantly broader features
indicative of larger expansion velocities.  We estimate the
photospheric velocity using the location of the \ion{Si}{2}$\lambda
6355$ absorption feature (Figure~\ref{fig:comp3}).  We find that the
for SN\,2009nz the minimum is located at $\lambda_{\rm rest}\approx
6000$ \AA\ (with an uncertainty of about 50 \AA), corresponding to a
velocity of about $v_{\rm ph}\approx 17,000$ km s$^{-1}$ (with an
uncertainty of about 2500 km s$^{-1}$).  A comparison to the inferred
\ion{Si}{2}$\lambda 6355$ velocities of the previous GRB-SNe
\citep{cbl+10} indeed indicates a similar velocity to SN\,2006aj
($\approx 19,000$ km s$^{-1}$), and lower velocity than SN\,1998bw
($\approx 24,000$ km s$^{-1}$) and SN\,2010bh ($\approx 25,000$ km
s$^{-1}$).

Using the photospheric velocity from our spectrum, along with the peak
brightness ($M_V$) and width of the photometric light curve
(parametrized as $\Delta m_{15}$), we can determine the basic
explosion properties of SN\,2009nz, namely the kinetic energy ($E_K$),
the ejecta mass ($M_{\rm ej}$), and the mass of synthesized $^{56}{\rm
Ni}$ ($M_{\rm Ni}$).  The light curve presented in \citet{cbp+10}
indicates $M_V=-19.0\pm 0.2$ mag and $\Delta m_{15}(V)=1.3\pm 0.3$
mag.  Using the formulation of \citet{dsg+10}, based on the
theoretical model of \citet{arn82}, we
infer\footnotemark\footnotetext{We use a mean difference of $\approx
0.3$ mag between $\Delta m_{15}(V)$ and $\Delta m_{15}(R)$, as well as
between $M_V$ and $M_R$, as found by \citet{dsg+10}, to place
SN\,2009nz on their grid of models.  The models assume energy
deposition from $^{56}{\rm Co}$ and $^{56}{\rm Ni}$, a homogeneous
density distribution of the ejecta, and a fixed optical opacity.}
$M_{\rm Ni,\odot}\approx 0.35$ and $M_{\rm ej,\odot}^{3/4}/
E_{K,51}^{1/4}\approx 1.0$ (here $M_{\rm ej,\odot}$ is the ejecta mass
in units of solar masses and $E_{K,51}$ is the kinetic energy in units
of $10^{51}$ erg).  From our inferred velocity we also find
$E_{K,51}/M_{\rm ej,\odot}\approx 1.7$, and therefore $E_{K,51}\approx
2.3$ and $M_{\rm ej,\odot}\approx 1.4$.  As with the overall shape of
the spectrum, the inferred explosions properties of SN\,2009nz most
closely resemble those of SN\,2006aj, for which \citet{mdn+06}
estimated $M_{\rm Ni,\odot}\approx 0.2$, $E_{K,51}\approx 2$, and
$M_{\rm ej,\odot}\approx 2$.  On the other hand, they are
significantly lower than for SNe 1998bw and 2003lw; see
Table~\ref{tab:exp} and Figure~\ref{fig:exp_prop}.  The maximal values
of the explosion properties given the uncertainties in $M_V$, $\Delta
m_{15}$, and $v_{\rm ej}$ are: $M_{\rm Ni,\odot}\approx 0.6$,
$E_{K,51}\approx 8.4$ and $M_{\rm ej,\odot}\approx 3.5$.

\section{Discussion and Conclusions}
\label{sec:conc}

We presented a spectrum of the photometric bump (designated
SN\,2009nz) associated with the standard cosmological \grb, and
unambiguously demonstrated that it is a broad-lined SN, similar to the
Type Ic SNe associated with nearby GRBs.  The best match over the
wavelength range of our spectrum ($\lambda_{\rm rest}\approx
4000-6300$ \AA) is provided by SN\,2006aj; SNe 1998bw and 2010bh, on
the other hand, exhibit broader features indicative of larger
expansion velocities.  Identifying the absorption feature centered at
$6000$ \AA\ with \ion{Si}{2}$\lambda 6355$, we infer a photospheric
velocity about $17,000$ km s$^{-1}$ at $\delta t_{\rm rest}\approx
16.3$ d, similar to that of SN\,2006aj, and lower than SNe 1998bw and
2010bh.  Combined with the SN light curve properties, we find that the
explosion properties of SN\,2009nz are $E_{K,51}\approx 2.3$, $M_{\rm
ej,\odot}\approx 1.4$, and $M_{\rm Ni,\odot}\approx 0.35$, similar to
those of SN\,2006aj.

These results demonstrate that beyond the basic need to
spectroscopically confirm photometric bumps as supernovae,
spectroscopy of GRB-SNe at $z\gtrsim 0.3$ provides two key
measurements.  First, the spectra allow us to determine the SN type.
To date, all the nearby well-studied GRB-SNe have been classified as
broad-lined Type Ic events.  At $z\gtrsim 0.3$ there are claims for a
normal Type Ic event (GRB\,021211/SN\,2002lt; \citealt{dmb+03}) and a
Type IIn event (GRB\,011121/SN\,2001ke; \citealt{gsw+03}).  However,
both of these claims are problematic: SN\,2002lt equally resembles the
peculiar Type Ia SN\,1991bg, while SN\,2001ke suffers from large, and
uncertain, extinction leading to an uncertain spectral shape.  In
addition, in neither case was the host contribution accounted for,
leading to possible contamination of the spectra.  The remaining three
events at $z\gtrsim 0.3$ with spectroscopic observations (XRF\,020903,
GRB\,050525A, and GRB\,101219B) broadly resemble SN\,1998bw
\citep{skf+05,dmb+06,ssf+11}, but no attempt has been made to extract
their physical properties.  Here, we find that SN\,2009nz is also a
broad-lined event, but for the first time have a sufficiently large
comparison sample to demonstrate a closer similarity to SN\,2006aj
than to the canonical SN\,1998bw.

\begin{figure}
\epsscale{1}
\plotone{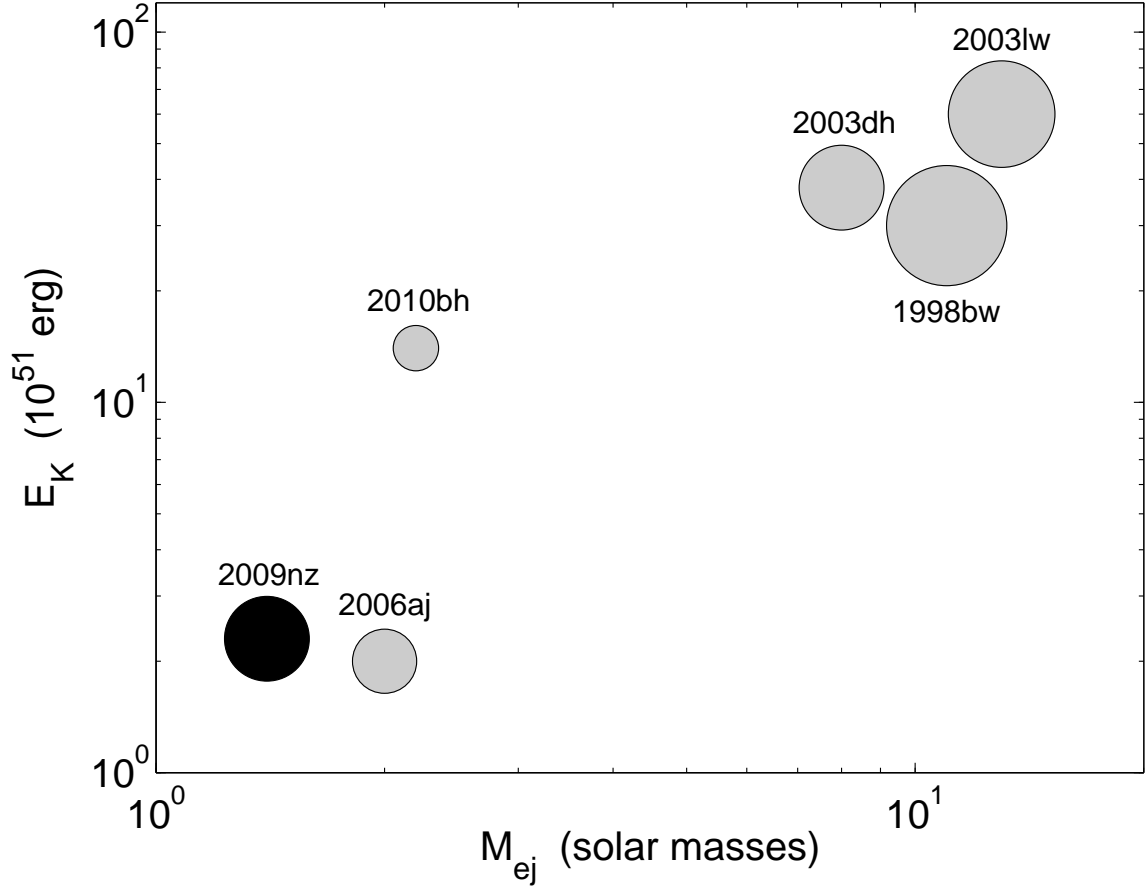}
\caption{Explosion properties of the well-studied nearby GRB-SNe (gray
circles), along with our inferred properties of SN\,2009nz (black
circle).  The area of the symbols is linearly proportional to the mass
of synthesized $^{56}{\rm Ni}$ for each event.  There appears to be a
broad correlation between the three explosion properties, with the
most energetic GRB-SNe producing the largest $^{56}{\rm Ni}$ and
ejecta masses.  The overall correlation between $E_K$ and $M_{\rm ej}$
reflects the range of ejecta velocities of $\sim 10,000-30,000$ km
s$^{-1}$.  References for the individual events are given in
Table~\ref{tab:exp}.
\label{fig:exp_prop}} 
\end{figure}

Second, the spectra allow us to measure the photospheric velocity and,
along with the light curve properties, to infer the basic explosion
parameters (the light curve by itself only provides a ratio of the
kinetic energy and ejecta mass).  Thus, spectroscopy provides deeper
insight into the physical properties of GRB-SNe than light curves
alone.  Here, for the first time, we measured the expansion velocity
and explosion properties of a GRB-SN at $z\gtrsim 0.2$.  Using these
measurements we find that SN\,2009nz follows the overall trend of
lower $M_{\rm Ni}$ for events with lower $E_{K}$ and $M_{\rm ej}$ seen
in the nearby GRB-SNe (Figure~\ref{fig:exp_prop}).

In the same vein, the similarity in spectral and explosion properties
of SN\,2009nz and SN\,2006aj is instructive in light of the
differences between the prompt emission properties of GRB\,091127 and
XRF\,060218.  The former is a typical cosmological GRB with $E_{\rm
\gamma,iso}\approx 1.1\times 10^{52}$ erg, a duration of
$T_{90}\approx 9$ s, and a rest-frame $E_p\approx 25$ keV
\citep{gcn10204}, while the latter is an X-ray flash with $E_{\rm
\gamma,iso}\approx 6.2\times 10^{49}$ erg, a duration of
$T_{90}\approx 2100$ s, and a rest-frame $E_p\approx 5$ keV
\citep{cmb+06}.  These differences provide evidence for a lack of
correlation between the prompt emission and associated SN properties.
Moreover, \citet{mdn+06} suggested that XRF\,060218/SN\,2006aj was
powered by a neutron star instead of a black hole based on the lower
ejecta mass and kinetic energy compared to SNe 1998bw, 2003dh, and
2003lw, as well as the low energy of the prompt emission.  However,
GRB\,091127 is typical of the cosmological GRB population, which is
thought to be powered by black holes \citep{mw99}, and yet its
associated SN\,2009nz shares the properties of SN\,2006aj.  This
either casts doubt on the idea that the properties of SN\,2006aj (or
of GRB-SNe in general) require a neutron star central engine, or
alternatively suggests that standard GRBs can be powered by either
neutron stars or black holes, but with no clear imprint on the prompt
emission properties.

From an observational point of view, we stress that while the broad SN
features are identifiable in the bump (supernova+host) spectrum of
SN\,2009nz, subtraction of the host spectrum provides a more reliable
and robust identification of the SN features.  Accounting for the host
contribution in GRB-SN spectra at $z\gtrsim 0.3$ is critical since in
all cases to date the host galaxy contributes about half of the
observed flux\footnotemark\footnotetext{The afterglow contribution, on
the other hand, is generally $\lesssim 10\%$, and it is also less of a
concern due to its featureless spectrum.}.  None of the studies
published to date accounted for the actual host spectrum, either
neglecting its contribution altogether (GRB\,021211: \citealt{dmb+03};
GRB\,011121: \citealt{gsw+03}), assuming a template blue galaxy
spectrum (GRB\,050525A: \citealt{dmb+06}; XRF\,020903:
\citealt{skf+05}), or assuming a flat spectrum (GRB\,101219B:
\citealt{ssf+11}).  In this study we find that either an actual
spectrum of the host or an appropriate template matched to the host
colors should be subtracted.

We end by noting that the small sample of GRB-SNe with detailed
spectroscopic observations already reveals a wide dispersion in
explosion properties, with about a factor of 30 in $E_K$, a factor of
15 in $M_{\rm ej}$, and a factor of 7 in $M_{\rm Ni}$
(Table~\ref{tab:exp} and Figure~\ref{fig:exp_prop}).  This also seems
to be the case for the two events associated with standard GRBs ---
GRB\,030329/SN\,2003dh and GRB\,091127/SN\,2009nz.  The wide
dispersion contradicts earlier claims for a narrow range of GRB-SN
properties, mainly their kinetic energies \citep{krg+07}.  Clearly, a
larger number of events is required to fully sample the range of
explosion parameters, to evaluate whether there really exists a
dispersion in GRB-SN types, and to investigate any correlations with
the GRB properties or differences between the SNe associated with
sub-energetic and standard GRBs.  Such a study requires a concerted
effort to determine the afterglow properties from broad-band modeling,
to measure the SN light curves, and to obtain spectra of the bumps and
host galaxies.  With a {\it Swift} detection rate of $2-3$ $z\sim 0.5$
GRBs per year such a concerted effort is both desirable and
achievable.

\acknowledgements E.B.~acknowledges partial support from Swift AO6
grant NNX10AI24G.  Based in part on observations obtained at the
Gemini Observatory, which is operated by the Association of
Universities for Research in Astronomy, Inc., under a cooperative
agreement with the NSF on behalf of the Gemini partnership: the
National Science Foundation (United States), the Science and
Technology Facilities Council (United Kingdom), the National Research
Council (Canada), CONICYT (Chile), the Australian Research Council
(Australia), Minist\'{e}rio da Ci\^{e}ncia e Tecnologia (Brazil) and
Ministerio de Ciencia, Tecnolog\'{i}a e Innovaci\'{o}n Productiva
(Argentina).  This paper includes data gathered with the 6.5 meter
Magellan Telescopes located at Las Campanas Observatory, Chile.


\clearpage
\begin{deluxetable}{cccc}
\tabletypesize{\footnotesize}
\tablecolumns{4} 
\tabcolsep0.15in\footnotesize
\tablewidth{0pt} 
\tablecaption{Magellan/IMACS Host Galaxy Photometry
\label{tab:photom}}
\tablehead{
\colhead{Filter}               &
\colhead{Exposure Time}        &
\colhead{$m_{\rm AB}$}         &
\colhead{$F_\nu\,^a$}          \\
\colhead{}                     &              
\colhead{(s)}                  &            
\colhead{(mag)}                &            
\colhead{($\mu$Jy)}           
}            
\startdata
$g$ & $420$ & $24.05\pm 0.14$ & $0.99\pm 0.14$ \\ 
$r$ & $720$ & $23.34\pm 0.05$ & $1.84\pm 0.09$ \\
$i$ & $480$ & $22.89\pm 0.07$ & $2.70\pm 0.19$ \\
$z$ & $720$ & $22.57\pm 0.10$ & $3.56\pm 0.36$
\enddata
\tablecomments{$^a$ Corrected for Galactic extinction of
$E(B-V)=0.038$ mag \citep{sfd98}.}
\end{deluxetable}

\clearpage
\begin{deluxetable}{ccccl}
\tabletypesize{\footnotesize}
\tablecolumns{5} 
\tabcolsep0.25in
\tablewidth{0pt} 
\tablecaption{Explosion Properties of GRB-SNe
\label{tab:exp}}
\tablehead{
\colhead{GRB-SN}               &
\colhead{$M_{\rm Ni}$}         &
\colhead{$E_{K}$}              &
\colhead{$M_{\rm ej}$}         &
\colhead{Reference}                 \\
\colhead{}                     &              
\colhead{(M$_\odot$)}          &            
\colhead{($10^{51}$ erg)}      &            
\colhead{(M$_\odot$)}          & 
\colhead{}                                   
}            
\startdata
1998bw & 0.7  & 30  & 11  & \citet{imn+98} \\ 
2003dh & 0.35 & 38  & 8   & \citet{mdt+03} \\
2003lw & 0.55 & 60  & 13  & \citet{mdp+06} \\
2006aj & 0.2  & 2   & 2   & \citet{mdn+06} \\
2010bh & 0.1  & 14  & 2.2 & \citet{cbg+11} \\
2009nz & 0.35 & 2.3 & 1.4 & This paper
\enddata
\tablecomments{}
\end{deluxetable}

\clearpage

\clearpage

\clearpage

\clearpage

\end{document}